\documentclass[preprint]{aastex}
\shorttitle{The Lupus Transit Survey}
\shortauthors{Bayliss et al.}

\begin{document}
\title{The Lupus Transit Survey for Hot Jupiters: \\Results and Lessons}
\author{Daniel D R Bayliss} 
\affil{Research School of Astronomy and
Astrophysics, The Australian National University, Mount Stromlo
Observatory, Cotter Road, Weston Creek, ACT 2611, Australia}
\email{daniel@mso.anu.edu.au}
\author{David T F Weldrake} 
\affil{Harvard-Smithsonian Center for Astrophysics,
60 Garden St MS-51, Cambridge, MA 02138, USA}
\email{dweldrak@cfa.harvard.edu}
\author{Penny D Sackett} 
\affil{Research School of Astronomy and
Astrophysics, The Australian National University, Mount Stromlo
Observatory, Cotter Road, Weston Creek, ACT 2611, Australia}
\email{penny.sackett@anu.edu.au}
\author{Brandon W Tingley} \affil{Instituto de Astrof\'{i}sica de Canarias
  C/ V\'{i}a L\'{a}ctea, s/n E38205 - La Laguna (Tenerife), Spain}
\email{btingley@iac.es}
\author{Karen M Lewis} \affil{School of Mathematical Sciences, Monash
University, Clayton, Victoria 3800, Australia}
\email{karen.lewis@sci.monash.edu.au}

\begin{abstract}
We present the results of a deep, wide-field transit survey targeting
``Hot Jupiter'' planets in the Lupus region of the Galactic plane
conducted over 53 nights concentrated in two epochs separated by a
year.  Using the Australian National University 40-inch telescope at Siding Spring
Observatory (SSO), the survey covered a 0.66~deg$^{2}$ region close to the
Galactic Plane ($b=11^{\circ}$) and monitored a total of 110,372
stars ($15.0<V<22.0$).  Using difference imaging photometry,
16,134 light curves with a photometric precision of $\sigma<0.025$~mag
were obtained. These light curves were searched for transits, and four
candidates were detected that displayed low-amplitude variability
consistent with a transiting giant planet.  Further investigations,
including spectral typing and radial velocity measurements for some
candidates, revealed that of the four, one is a true planetary
companion (Lupus-TR-3), two are blended systems
(Lupus-TR-1 and 4), and one is a binary (Lupus-TR-2).  The results of this successful survey are instructive for optimizing the
observational strategy and follow-up procedure for deep searches for
transiting planets, including an upcoming
survey using the SkyMapper telescope at SSO. 

\end{abstract}


\section{Introduction}
To date, approximately 15\% of all known extrasolar planets have been
discovered by detecting the photometric signal produced when the
planet transits its host star.\footnote{http://exoplanet.eu/} This
transit technique offers not only a fruitful discovery mechanism, but
also a means of determining the planetary radius, and with
spectroscopic follow up, true planetary mass and density.
Furthermore, for bright stars, the atmospheres of transiting planets
can be studied, both in emission \citep{charbonneau05,deming05} and
absorption \citep{vidal04,tinetti07}.

Wide-field transit surveys generally target bright ($V<12$) stars, and
have yielded the majority of transiting planet discoveries (for
example, HATNet, \citet{bakos07}; TrES, \citet{alonso04}; WASP, \citet
{cameron07}; and XO, \citet{mccullough06}).  Deeper, narrower surveys are less
numerous in the literature, yet have also been successful, most
notably the OGLE Survey \citep{udalski02}, which has detected
several confirmed transiting planets down to $V\sim17.0$
\citep{udalski07}.

Deep surveys require larger aperture
telescopes (typically $\sim1$~m), and the faint candidates they produce
are difficult to follow up.  However, for statistical studies of
extrasolar planet populations, deep surveys will grow in
importance.  Due to the advent of very large format, high-quality
mosaic CCD cameras, the huge number of main sequence dwarfs in each
field, and the improvements in follow-up techniques, large numbers of
planets can be found.  Deep surveys
also monitor a much higher fraction of late type stars, about which
less is known in terms of planetary companions.
  
We report here on the strategy and results from the Lupus Transit
Survey, a deep survey focusing on monitoring stars in the magnitude range
$14.5<V<19.5$ in a 0.66 deg$^2$ field. More than 16,000 stars were
monitored with a $1 \rm m$ class
telescope (coupled with a large format CCD detector) with sufficient
precision ($\sigma<0.025$), cadence (6~minutes), and survey duration
(26 contiguous nights in 2005, 27 contiguous nights in 2006) to have high sensitivity to
detect transiting Hot Jupiters out to a few days in orbital period.

The motivation for the Lupus Transit Survey was threefold. First, the
survey is a deep transit search in its own right, capable of
discovering interesting planetary systems in a previously unsampled
part of the sky.  Second, the survey acts as a control field for the
previous globular cluster transit surveys of
\citet{weldrake05,weldrake07}, both of which produced significant null
results using the same telescope, instrumental setup, and a very
similar observational strategy.  Third, the Lupus Survey acts as an
excellent pilot study to refine further the techniques and strategies
for the future 5.7~deg$^{2}$ SkyMapper Transit Survey
\citep{bayliss07}, currently being undertaken from the same site.

We describe the observations and data reduction for the survey in
Section \ref{observations}.  In Section \ref{CMDandAstrom} we set out
the astrometric and photometric properties of the Lupus field.
Section \ref{photometry} describes the technique used to produce
precise relative photometry for stars in the survey field.  Section
\ref{detection} outlines the analysis of the light curves to search for
transiting planets and the procedure used to cull false candidates
based on the survey data.  Section \ref{fitting} outlines how we
fitted model transits to the photometry of candidates to determine the
depth, duration, and center times for each candidate.  Section \ref{typing} describes the follow-up
spectral typing that was undertaken on promising candidates, while
Section \ref{RV} describes the radial velocity measurements obtained
for three candidates. Section \ref{2006} describes the importance of
the extra data from 2006 in determining the candidates' nature.
Section \ref{y_band} sets out the good seeing, near-infrared (near-IR) imaging we used to look
for blended stars near our candidates.  Our
conclusions and discussion are set out in Section \ref{conclusions}.

Our Lupus Survey detected the Hot Jupiter Lupus-TR-3b, which is described
fully in \citet{weldrake08b}; only an overview is included
in this paper. The survey also uncovered 494 previously unknown
variable stars, which are published in a second companion paper:
\citet{weldrake08}.  A project to expand this survey, SuperLupus, is
currently underway to search for longer period
planets\citep{bayliss08}.  A full statistical analysis for
planet frequency in this field will be undertaken at the
conclusion of the SuperLupus project.  This paper, therefore,
refrains from any statistical analysis of the planet frequency.

\section{Observations and Data Reduction}
\label{observations}

\subsection{The Instrumental Set-Up} 
The Lupus Transit Survey was undertaken using the Wide Field Imager
(WFI) on the Australian National University  (ANU) 40-inch telescope at Siding Spring Observatory (SSO)
in New South Wales, Australia.  Full details of the telescope and WFI
are set out in Table \ref{details}. Although the detector was
52$\arcmin$ on each side, throughout the survey one of the eight CCDs was
not functional, reducing the area of sky monitored to 0.66~deg$^{2}$.
A single broadband $V+R$ filter was used, covering the combined
wavelength range of Cousins $V$ and $R$, increasing the resulting
signal-to-noise ratio (S/N) of the photometry for a fixed integration
time. This same telescope and instrument configuration has been successfully
employed in the two previous transit surveys of globular clusters: 47
Tuc and $\omega$ Cen \citep{weldrake05,weldrake07}.

\subsection{Observational Strategy}    
The duration of a transit survey is critical to its prospects of
success.  Underestimating the required survey duration is certainly
one reason why transiting planet survey yields have been lower than
anticipated \citep{pont06}.  In order to mitigate against this effect,
a single field was observed intensely for two month-long periods
separated by a year.  An added advantage of this single-field strategy
was that no time was lost in moving the telescope between fields.
Such delays could have been lengthy given current pointing
difficulties with the ANU 40-inch telescope. A crowded field near the
Galactic plane was selected for the survey ($b=11^{\circ}$).  The exact survey field, centered at
R.A.~$=15^{\rm{h}}30^{\rm{m}}36.3^{\rm{s}}$,
decl.~$=$$-42^{\circ}53'53.0\arcsec$ (J2000), in the constellation
of Lupus, was chosen to maximize usable night-time hours, maintain a
reasonable minimum distance from the Moon, minimize the number of
bright (saturated) stars in the field, and minimize extinction from
interstellar dust \citep{schlegel98}.

An exposure time of $300$~s per image was adopted for all images.  As
the readout time of WFI is $50$~s, a cadence of approximately 6~minutes was achieved during the
observations.

\subsection{Data Acquisition and Reduction} 
A total of 2710 $V+R$ images of the survey field were obtained over 53
nights (26 nights in 2005 and 27 nights in 2006).  The
full width at half-maxima (FWHM) of the stellar point spread functions (PSFs)
were measured and recorded for each image.  
Due to the crowding in the field, only images with FWHM less than
2.2\arcsec\space were able to yield high-precision photometry, which left us with a total of 1783
images. Manual corrections to the pointing during the night
ensured that image shifts on the CCD were kept to within a few
pixels. This minimizes the effect of
pixel-to-pixel sensitivity variations in the resulting photometry, and
increases the number of usable sampled stars.

Data were reduced using standard
IRAF\footnote{IRAF is distributed by the National Optical Astronomy
Observatory, which is operated by the Association of Universities
for Research in Astronomy, Inc., under cooperative agreement with the
National Science Foundation.} tasks in the MSCRED package, whereby the images were trimmed and bias, dark and flat-field corrected using
calibration frames.  Biases were obtained daily, and sky-flats taken at twilight and dawn when weather
permitted.

\section{Colour-Magnitude Diagram and Astrometry}
\label{CMDandAstrom}
In addition to the survey images, seven 10~minute exposures of the
field were taken in  $V$, $R$ and $I$ bands during good seeing, photometric
conditions.  These deeper images were used for an accurate identification
of astrometric positions and colors for the stars in
the survey field.

A $V$, $V-I$ color-magnitude diagram (CMD) was produced from these images to place the detected transit candidates onto the standard
magnitude system (see Fig.\space\ref{CMD}).  The survey has a saturation limit at
$V\sim14.5$, and a faint limit of $V=22.0$ (19.4 for the transit search).  The DAOPHOT-derived errors and
calibration uncertainty (see the following paragraph) in our magnitude
determinations are also overplotted, along with the location of our
six transit candidates.

The CMD was calibrated via observations of MarkA standard stars
\citep{Lan1992,Stet2000} taken during the 2005 observing run in
photometric conditions.  More than four hundred standards were
identified via matching of astrometry\footnote{MarkA
astrometry and photometry downloaded from
http://www4.cadc-ccda.hia-iha.nrc-cnrc.gc.ca/community/STETSON/standards/}. The
mean and standard deviation of the magnitude shift was determined for each
CCD and each filter independently. The resulting calibration
uncertainty is 0.03 mag in $V$ and 0.05 mag in $V-I$. As an
additional check, the CMD was also cross-correlated with 2000 stars
within 20$'$ of the Lupus field center in the NOMAD online
catalog\footnote{http://www.nofs.navy.mil/data/fchpix/} and found to
match within the precision of the catalog.

The astrometry for all sampled stars was determined via a search of
the USNO CCD Astrograph Catalog (UCAC1), cross-identifying astrometric
standard stars within the field. Several hundred such stars per CCD were
successfully identified, producing an accurate determination of the
astrometric solution for the stars in each CCD independently; the
resulting astrometric calibration accuracy was 0.25$''$. The layout of the field,
with the astrometry of all the stars, can be seen in our companion
paper \citep{weldrake08}.

\section{Photometry}
\label{photometry}
An application of difference imaging analysis (DIA), originally described as
the optimal PSF-matching package of \citet{alard98}, was used to produce
precise time-series photometry for each star.  A modified form
of this package was used \citep{wozniak00}.

DIA matches the stellar PSF throughout a large database of images,
thereby dramatically reducing the systematic effects from varying
atmospheric conditions on the output photometric precision. This
method allows ground-based observations the best prospects of
detecting small-amplitude brightness variations in faint targets. DIA
is an excellent method to obtain photometric time series in crowded
fields as a large number of pixels are used to characterize PSF
differences, improving the PSF-matching process.

Initial stellar flux measurements via DIA are made from profile-fitted
photometry on a template frame, produced by median-combining a number
of the best-quality images with small offsets, in this case, 74 images
with offsets $\le$20 pixels (8$\arcsec$ on the sky or 2$\%$ the width of
a CCD). Flux measurements on this template are used as the zero
point of the resulting differential stellar time series for each
individual star.

Stellar positions were found on a best-seeing reference image, and all
the subsequent images, including the template, were then registered to
these positions.
The best PSF-matching kernel was determined and each registered image
was subsequently subtracted from the template. The residuals from this
method are dominated by photon noise.  Any object that
changed brightness between the image in question and the template was recorded as
a bright or dark spot in the residual map.

This method of differential photometry produces a time series in
differential flux counts rather than in the usual magnitude units.  To
convert to a standard magnitude system, we measured the total counts
for each star on the template frame using the DAOPHOT/PSF package in
IRAF, with the parameters set to the same values used in the
DIA routine. We then converted the time-series photometry into
magnitude units by the relation
\begin{eqnarray}
\Delta m_{i}=-2.5log[(N_{i}+N_{ref,i})/N_{ref,i}], \nonumber
\end{eqnarray}

where $N_{{ref,i}}$ is the total flux of star ${i}$ on the
template image, and $N_{{i}}$ is the original difference flux in the
time-series as produced with the photometry code.  It is important to
note when combining differential fluxes with
DAOPHOT-derived reference image photometry, we must correct for errors
based on the individual apertures used, as described in Appendix B of
\citet{H2004}. An aperture correction was used to determine the
scaling between the two fluxes, which was performed on the stellar
DAOPHOT magnitudes. Our PSF magnitudes were consistently 0.06~mag brighter than the aperture-derived values (using the same
aperture values as in the differential photometry). We hence shifted
our magnitude zero point to $25.0-0.06=24.94$. This correction
ensures that the differential imagery output is accurately represented
in magnitude units.

Since the Lupus Transit Survey data were obtained over 53 nights in
classical observing mode, our observations span a wide range of airmasses,
weather conditions, and phases of the Moon.  All of these effects
produce systematic trends in the light curves, along with other higher
order systematic effects contributing to the so-called ``red''
noise. The SYSREM algorithm \citep{tamuz05} was used in order to
reduce these systematic trends in our data set. This principle-component technique algorithm is
often used in large transit surveys to remove effects in the
light curves that may be
common to a large data set of stars.  Using only nonvariable stars
in the algorithm, and running 10 iterations, an average improvement of
$\sim50\%$ in the photometric precision was achieved.  The improvement
was greatest for the bright stars, whose photometry is dominated
by systematics as opposed to faint stars whose scatter is dominated
by photon noise.  This is significant, since the survey is most sensitive
to detecting planet transits against bright stars.

A total of 16,134 light curves in the field displayed post-SYSREM
standard deviations (rms) of less than 0.025 mag, and it was these
light curves we analyzed for transiting planets.  The total data set of
110,372 light curves was analyzed only for large-scale variability. A
total of 494
previously unknown variable stars were identified in the field; these
have been published by \citet{weldrake08}. 

Figure \ref{RMS} shows the photometric precision achieved for those
stars with $15.0<V<19.4$, which includes the 16,134 stars
analyzed for transiting planets.  The overplotted solid line shows the
total of the theoretical photon noise for the star and sky, which
describes the photometry well.

\section{Transit Search and Candidate Selection}
\label{detection}
Since the primary goal for the Lupus Survey was the discovery of transiting
Hot Jupiters, immediately after the 2005 observing run, we searched
for transitlike events in the 2005 data set.  As the field was only
visible for follow-up work for $\sim$three months after the
observing run, it was important to identify candidates as soon
as was possible.  After the 2006 observing run, the combined data set
was searched again; the results of this are
discussed in Section \ref{2006}.  

The Box-fitting Least-Squares (BLS) algorithm
\citep{kovacs02} was used to search for transit signals in our light curves.
The BLS algorithm searches for the first-order shape of a transit via least
squares fitting of step functions to the phase-wrapped data, and is
widely used in transit surveys. 

For each light curve, BLS was used to compute the best period and a
corresponding significance of that period, the so-called Signal Detection
Efficiency (SDE).  Light curves were ranked according to their SDE, and
all were examined wrapped at the BLS determined best-fit period and
integer aliases thereof. 

From the 2005 data, four candidates were identified with high SDE and
phase-wrapped light curves consistent with that expected for a transiting Hot Jupiter (see
Fig.\space\ref{candsplot}).  Two further candidates, Lupus-TR-5 and
Lupus-TR-6 showed transitlike signals, (see Fig.\space\ref{candsplot}),
however inspecting the raw images revealed that both were on, or very
close to, defective columns on the CCD.  We conclude that the low level
signal we detected in these candidates is due to systematic effects as
they drift over the defective CCD column during the course of the
observations, and we do not discuss these candidates further.

The four promising candidates and
their properties (R.A., decl., $V$ mag and $V-I$
color) are set out in Table \ref{cand_data}.  Figure \ref{candsplot} shows the
phase-wrapped $V+R$ photometry (with open circles for the 2005 data, filled circles for
2006). The best-fitting transit model (see Section \ref{fitting}) is
also overplotted for each. Figure \ref{TR12} similarly shows the
photometry for the time of anti-transit for Lupus-TR-1 and Lupus-TR-2,
described further in Section \ref{2006}.
 
For each candidate the total signal to noise for the transit feature was
 calculated using the equation:
\begin{eqnarray}
 S/N= d \sqrt{{n_{t}} / \sigma}, \nonumber
\end{eqnarray}
where $\sigma$ is the out-of-transit rms for the lightcurve, d
is the transit depth (both in magnitude units), and $n_{t}$ is the
number of in-transit points in the light curve.  In addition, the number
of observed transits for each candidate was determined as a measure of
robustness.

The $\eta$ diagnostic \citep{tingley05} was calculated for
each candidate to help weed out the most likely astrophysical false
positives.  The diagnostic $\eta$ is a measure
of how consistent the transiting candidate's depth, period, and duration
are with the expected theoretical values for a transiting extrasolar
planet.

The results for the S/N, number of observed transits, and
$\eta$ diagnostic are set out in Table \ref{cand_properties}.  By way of
comparison, the Lupus candidate diagnostic numbers are plotted with
those of the confirmed OGLE planets (as of 2007) in Figure \ref{eta}.  We
note that Lupus-TR-1 and TR-4 have $\eta$ higher than most of the OGLE
planets, increasing the likelihood that they are
false positives. Lupus-TR-3 alone has an $\eta$ value
comparable to the OGLE planets.

Careful examination of the images allowed us to show that Lupus-TR-4
was probably not planetary in origin without the need for follow-up observations.
Lupus-TR-4 had a close neighbor, which had already been
identified as
an eclipsing binary system as part of our search for
variable stars in the field \citep{weldrake08}.  Upon inspection it
was found that the period and phase of Lupus-TR-4 matched that of its
eclipsing binary neighbor, although with a far lower amplitude, which
gave it a depth consistent with a planetary transit. We
conclude that the signal seen in Lupus-TR-4 is the result of blending
with its neighbor.  This still left three promising candidates
(Lupus-TR-1, TR-2, and TR-3) which warranted further follow-up work.

\section{Transit Fitting}
\label{fitting}
To determine the parameters of the detected candidates, a transit fitting algorithm
was developed. This algorithm uses the transit models of
\citet{MA2002} to determine the best-fitting set of system parameters
for a fixed orbital period (and quadratic limb-darkening coefficients
appropriate for the spectral type of the primary), using a
minimization of the rms between the data and the model as a
fit statistic.

For each candidate, a large number of
transit models are produced ($\sim$100,000) for various stellar radii,
orbital inclinations and planet radii for a fixed orbital period and
stellar mass appropriate to the respective spectral type (see Section \ref{typing}). The orbital
semimajor axis is determined from these parameters, with the exact
range of model parameters dependent on the grazing/equatorial
appearance of the transit itself. Each individual model is then
compared to the phase-wrapped photometry (in units of hours from
transit center time) and a resulting rms residual is
obtained. The model with the lowest residual is taken as the
best-fitting description of the data.  The fits are displayed as solid
lines overplotted on the photometry in
Figure \ref{candsplot}.  From this best fit, we derive the  maximum transit depth ($d$) and the duration
for transit ($T_{14}$) for each candidate (see Table \ref{cand_properties}).

Uncertainties, also seen in Table \ref{cand_properties} were
determined via bootstrap resampling.  A variation on this fitting program was used to determine the transit
center times ($T_{C}$), also seen in Table \ref{cand_properties}.  Here we took the best-fitting
transit model for each candidate, comparing it to a well-sampled
single unphase-wrapped transit and determined the corresponding
transit center time that minimizes the residuals. Uncertainties were
again determined via bootstrap resampling.
 
\section{Spectral Typing}
\label{typing}
Spectra were obtained for the remaining candidates, Lupus-TR-1, TR-2,
and TR-3, using the Double Beam Spectrograph
(DBS) on the ANU 2.3~m telescope at SSO.  For Lupus-TR-1 and Lupus-TR-2
the 600-line blue grating was used, which gives resolution of 400 over a wavelength range
of 3600-4800~\AA\space(Fig.\space\ref{tr12spec}).  For Lupus-TR-3, a
300-line blue grating was used, yielding a resolution of 200 over
4000-6700~\AA\space(Fig.\space\ref{tr3spec}).

The spectra were reduced using standard IRAF tasks.  Corrections for
interstellar reddening assumed E(B-V)=0.182, an estimate based on dust
maps of the survey field \citep{schlegel98}. The reduced spectra were then
compared to the extensive MILES library of empirical spectra
\citep{sanchez04}, chosen as a comparison
set as it closely matched the spectral resolution we obtained for our
candidates.

Comparison to the MILES library indicated that both Lupus-TR-1 and
Lupus-TR-2 are early G-type dwarf stars, while Lupus-TR-3 is an early
K-type dwarf (see Figs.\space\ref{tr12spec}, \ref{tr3spec}, and Table
\ref{cand_data}). Since none of these three candidates are giant stars, we
can rule out the possibility
of a very large radius star being transited by a nonplanetary
secondary or a giant star blending a stellar eclipsing system as the
cause of the $\sim1\%$ dip.

\section{Radial Velocity measurements}
\label{RV}
RV measurements provide the ultimate verification as to
whether a candidate transit is caused by an orbiting planet by
determining the mass of the transiting object, which in turn defines
whether the object is stellar or planetary in nature. 

RV followup is a major hurdle for deep transit surveys,
and typically requires significant amounts of time on 8m class
telescopes to produce high-resolution spectra with sufficient
signal to noise for accurate RV cross-correlation. The
three promising Lupus candidates range from V=14.6 to V=17.4.  In an
effort to make best use of large aperture access, two different
instruments were used to obtain sufficiently precise RV measurements
for the determination of the nature of the transiting object.

\subsection{Radial Velocities from the AAT: Lupus-TR-1 and TR-2}
Lupus-TR-1 and Lupus-TR-2 are at the bright end of our survey
($V\approx14.5$), which meant that velocity measurements for these
candidates could be obtained using the 3.9m Anglo-Australian Telescope
(AAT) at SSO.  On 2006 July 19 and 20, we obtained data in service time
using UCLES, a high resolution echelle spectrograph at the AAT,
without the iodine cell.  The
data were taken with a central wavelength of 5414\AA \space and a
resolution of R$\approx$25,000.  The spectra cover the wavelength
range of 4450-7340\AA \space in 48 orders.  With 20~minute exposures,
S/N of 8 per pixel was achieved.

In addition, spectra were obtained for the bright RV
standard HR4695 on the same nights with the same setup.  Standard
tasks in the IRAF package ECHELLE were used
to reduce the spectra, and wavelength calibration was performed using
a number of ThAr arc lamp frames taken over the course of the
observations.  In total, 10 spectra were obtained for Lupus-TR-1 and
nine spectra for Lupus-TR-2.  We cross-correlated these spectra
against the HR4695 spectrum, selecting 40 orders with sufficient S/N
and a lack of strong telluric lines.

The RV for Lupus-TR-1 show no variation within the precision of our
measurements, as shown in Figure \ref{AAT}.  The service
time observing we acquired fell between the predicted times of maximum
and minimum RV amplitude, however on July 20 we obtained RV
measurements over approximately 4~hr, which is a significant
fraction of the period of Lupus-TR-1.  Given that the spectral typing of
this star suggests it is an early-type G-dwarf, then the lack of RV variations
over this period means that we can rule out a binary companion with a
mass greater than approximately 10$M_{J}$.  However, high-resolution
imaging with PANIC on  Magellan (see Section \ref{y_band}) indicates
this system is most likely a blend.  The final RVs for
Lupus-TR-1 are given in Table \ref{AAT_RV}.
  
The data for Lupus-TR-2 showed a discrete jump of 46 km\,s$^{-1}$ over
the course of just 80 minutes.  We believe that this must have been
due to a pointing error during the service observations, and therefore
the RV measurements for Lupus-TR-2 were discarded.
However further photometric monitoring of Lupus-TR-2 revealed it to be
a detached eclipsing binary (see Section \ref{2006}).

\subsection{Magellan Radial Velocity: Lupus-TR-3}
\label{magellan}
The most promising candidate, Lupus-TR-3, was too faint ($V=17.4$) for suitable
spectra to be obtained using UCLES on the AAT, and therefore
high-resolution spectra for its RV measurements were obtained
using the MIKE echelle spectrograph on Magellan II (Clay).  A signal
of $K=114\pm25$ m\,s$^{-1}$ was detected, which, along with extensive
blend modeling, indicated that Lupus-TR-3 is
transited by a Hot Jupiter with a mass of 0.81$\pm$0.18$M_J$. Transit
fitting provides a planetary radius of 0.89$\pm$0.07$R_{J}$, hence a
Jupiter-like density of 1.4$\pm$0.4~g\,cm$^{-3}$. These
observations and results are fully detailed in
\citet{weldrake08b}. This constitutes the faintest ground-based
detection of a transiting planet to date.

IR imaging of Lupus-TR-3 has very recently been obtained using
PANIC on Magellan, and these data are currently being analyzed using
a powerful image deconvolution technique (see \citet{sackett08}).   

\section{The Added Value of the 2006 Data}
\label{2006}
Lupus-TR-1, TR-2, and TR-3 were initially detected from the 2005 data
alone. 
In the case of Lupus-TR-2, the 2005 light curve did not contain any
data points at phase=0.5 (where phase=0 is the mid-point of the
transit).  However, the additional 27 nights of data taken
in 2006 May/June did cover this phase region, and revealed a
``secondary'' eclipse, which confirmed that Lupus-TR-2 was a detached
eclipsing binary
system (see Fig.\space\ref{TR12}) with twice the originally estimated
period.  The eclipse seen at phase 0.5 is
deeper than the original eclipse which was detected, meaning that we
had actually seen the shallow secondary transit in 2005, but by chance missed
the deep primary eclipse.

Lupus-TR-1 also exhibited a very shallow (0.8~mmag) secondary eclipse in the
combined 2005/2006 data set (see Fig.\space\ref{TR12}).  This is
indicative of a blended eclipsing binary, and this hypothesis was
subsequently examined with good-seeing, $Y$-band
imaging (see Section \ref{y_band}).

Lupus-TR-3 showed no secondary eclipse in the full 2005/2006 data set.
More transits of Lupus-TR-3 were observed in 2006 however, which
improved our knowledge of the period and phase of this planetary host.  

\section{$Y$ Band Imaging}
\label{y_band}
Good-seeing (0.65\arcsec), near-IR $Y$-band snapshots were obtained
using PANIC on Magellan I (Baade) to further help us determine the
nature of Lupus-TR-1 and further examine Lupus-TR-3.  This imaging of
Lupus-TR-1 revealed a neighbor at approximately 2.3$\arcsec$ from the
candidate (see Figure \ref{PANIC}).  Coupled with the 0.8mmag
secondary eclipse (see Fig.\space\ref{TR12}), no measured RV variations and a high $\eta$ diagnostic, we conclude that
this candidate is a blended eclipsing binary.

\section{Summary and Conclusions}
\label{conclusions}
We have performed a deep search for transiting short-period giant
planets in a single 0.66 deg$^{2}$ field above the Galactic
plane ($b\sim11^{\circ}$) using the ANU 40-inch telescope. Out of
16,134 stars with sufficient photometric precision ($\le$0.025 mag), we
identified four transit candidates (Lupus-TR-1 through TR-4, with $V=14.6
\rightarrow 17.4$). Without follow-up observations, we determined
that Lupus-TR-4 was a false positive as its light curve was
contaminated by a nearby variable.

Of the remaining three candidates, one (Lupus-TR-2) showed a deep
``secondary'' eclipse when we obtained further data in 2006, revealing
that the signal originally detected in the first year was in fact the secondary eclipse of a
detached eclipsing binary star.

Lupus-TR-1 showed no RV variations, which
coupled with the spectral typing of the primary (G1V) ruled out
a grazing binary system.  However, at a phase
of 0.5 the light curve did show the hint of a 0.8 mmag secondary
eclipse.  IR images from PANIC on Magellan revealed a relatively
bright neighbor, which strongly indicates this candidate is blended with an
eclipsing stellar binary that causes the transit signal. Lupus-TR-3
($V=17.4$) was found to be the host of a hot Jupiter planet, published
separately in \citet{weldrake08b}.

Candidates produced by this survey lie at the faint end of what is feasible for
obtaining precise RV follow up to confirm planets and
derive their mass.  The difficulties of follow up in this faint regime
are well outlined in \citet{pont08}, where a program to study OGLE
candidates is discussed.  Similar difficulties will also be experienced by
transiting planet surveys searching for low mass planets, such as
\textit{CoRoT} and \textit{Kelper}.  

In an attempt to minimize these difficulties, the upcoming SkyMapper
Transit Survey \citep{bayliss07} will be designed to monitor
stars brighter than $V\sim17$, with target numbers kept high by a combination of
SkyMapper's wide field of view and a multifield observational
strategy.

The Lupus Transit Survey benefited greatly from imaging the same field in 2005
and 2006, as this produced more complete phase coverage, particularly for
near integer-day periods preferentially found by ground-based surveys.
It also allowed us to more accurately determine
periods, which from the 2005 data alone were often confused by
aliasing.  This successful technique of several long, intense but
widely separated epochs will be adopted in the SkyMapper Transit Survey.

The good-seeing, near-IR imaging we obtained for candidates was a
relatively efficient method of checking for 
near neighbors that could be blended eclipsing binaries or
single stars that could cause an overestimation in the flux from the
primary star. Such imaging will form part of the SkyMapper Transit Survey
procedure of investigating the best candidates.

Lupus-TR-5 and Lupus-TR-6 showed promising transit-shaped dips in
their light curves, but were actually caused by a slight defect in the
CCD.  Mapping of any defects in the CCD will allow the early
identification of these types of impostors, and such mapping will be
implemented for the SkyMapper Transit Survey.  

During the follow up of candidates from the Lupus Transit Survey, we
utilized the facilities available at SSO, namely the ANU 2.3m telescope for spectral typing and the
AAT for medium-precision RV measurements
($\sim150$~m\,s$^{-1}$ on $V\sim15$ targets).  The success of these
facilities for this spectroscopic work indicates that they could be
important in follow up for other transit surveys in the Southern
Hemisphere.    

\acknowledgments
The authors thank Grant Kennedy for his assistance in obtaining the
survey data and Ken Freeman for his assistance in obtaining the spectrum
of Lupus-TR-3.  We acknowledge financial support from the Access to 
Major Research Facilities Programme, which is a component 
of the International Science Linkages Programme established 
under the Australian Government's innovation statement, 
Backing Australia's Ability.

{\it Facilities:} \facility{Magellan:Clay (MIKE)},
  \facility{Magellan:Baade (PANIC)}, \facility{SSO:1m (WFI)},
  \facility{ATT (DBS)}

\clearpage

\plotone{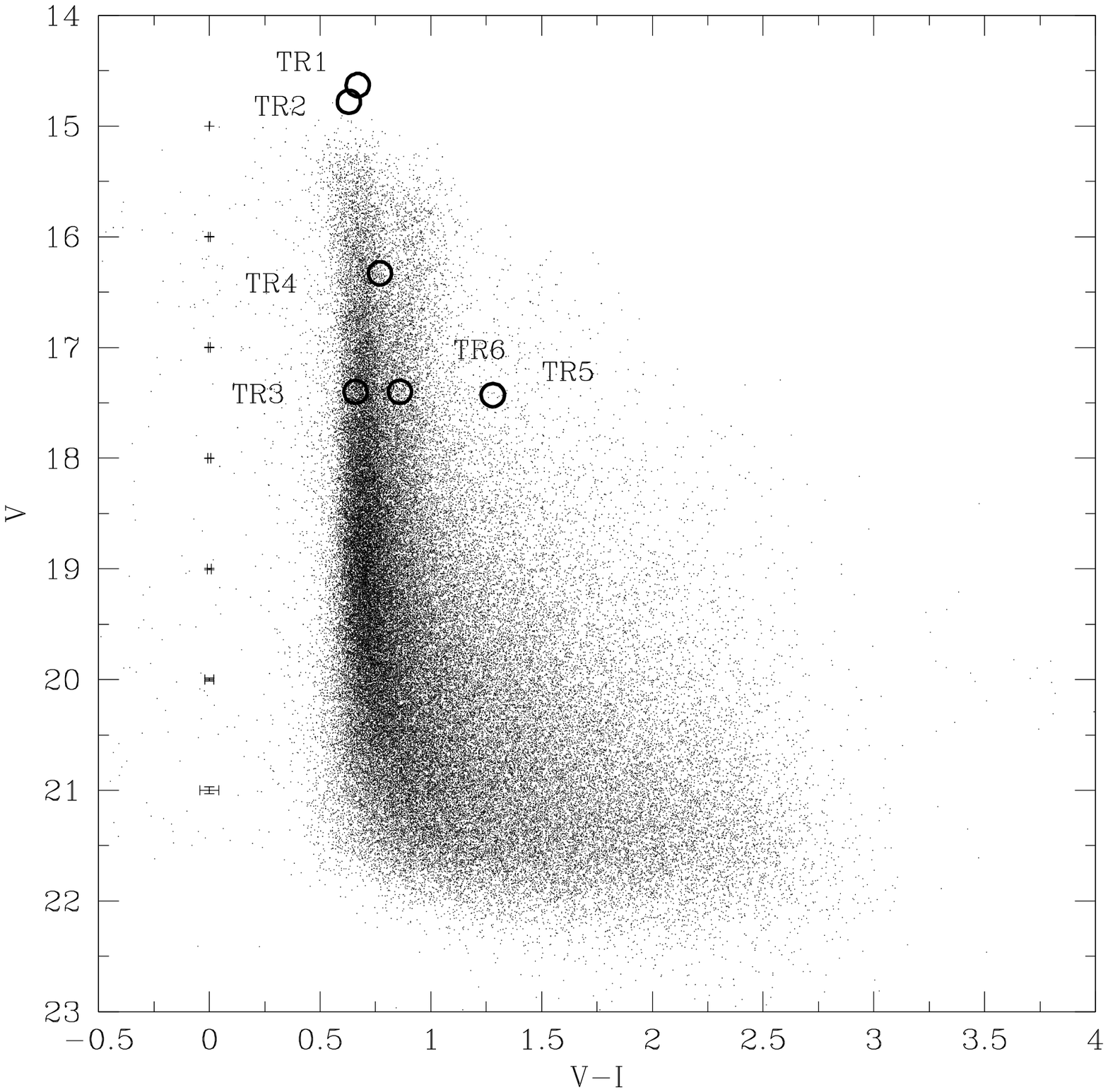} \figcaption[cmdplot.eps]{CMD for 95,358 stars in the Lupus Transit Survey field.  The
  overplotted error bars represent the output DAOPHOT errors in the
  magnitudes and the CMD calibration uncertainty added in
  quadrature. The positions of the six transiting planet candidates in
  Lupus are marked with open circles and labeled.\label{CMD}}

\plotone{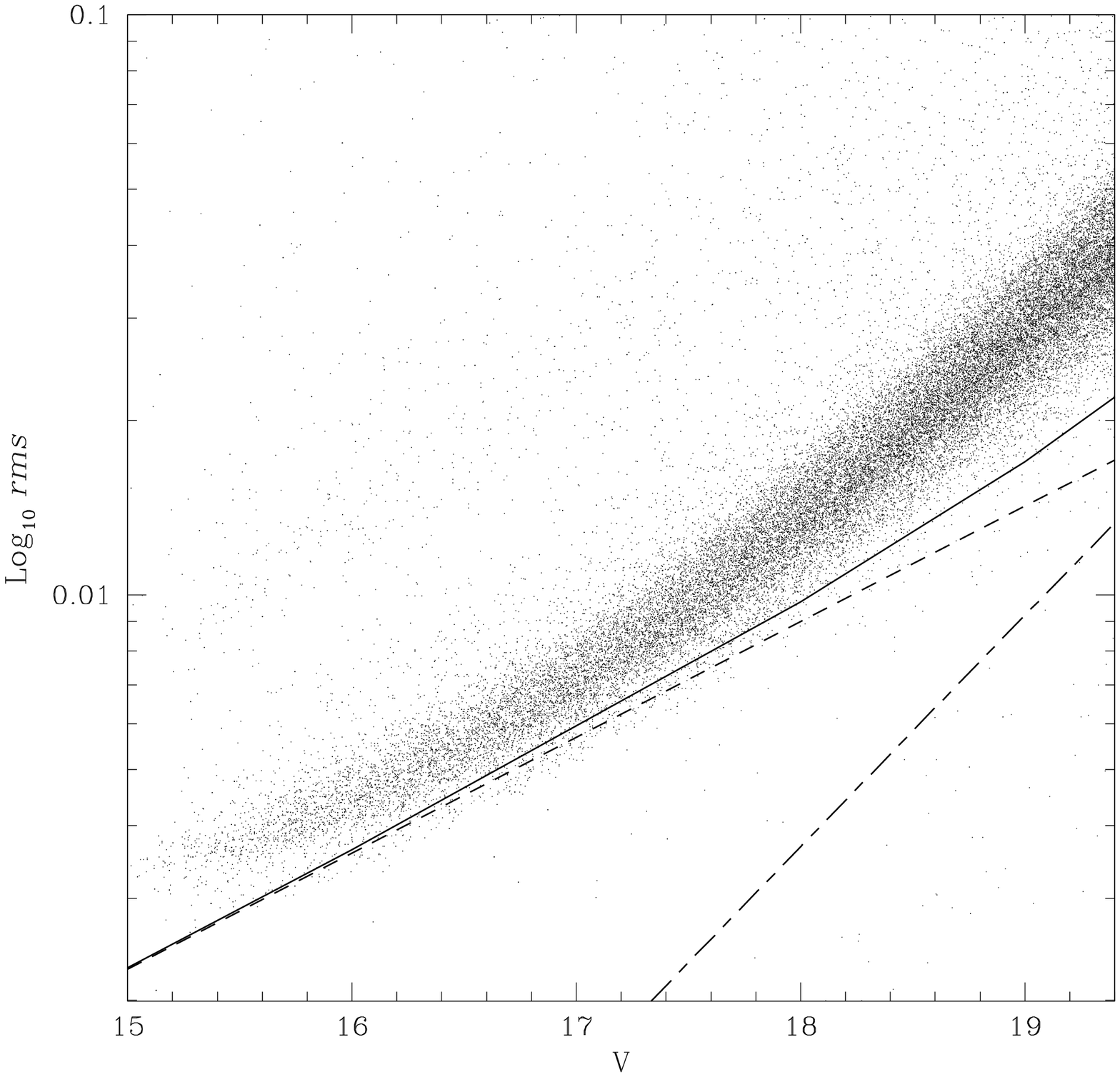} \figcaption[rmsplot.eps]{Photometric precision achieved for the stars monitored in the Lupus
  Transit Survey to a V magnitude limit of 19.38.  The dashed line
  represents the photon noise from the star.  The dash-dotted line
  represents photon noise from the sky background.  The solid line is
  the sum of these two lines (the total photon noise).  16,134 of these
  stars gave light curves with an (rms) of less than 0.025~mag, and these were the stars analyzed for transiting
  planets.\label{RMS}}

\plotone{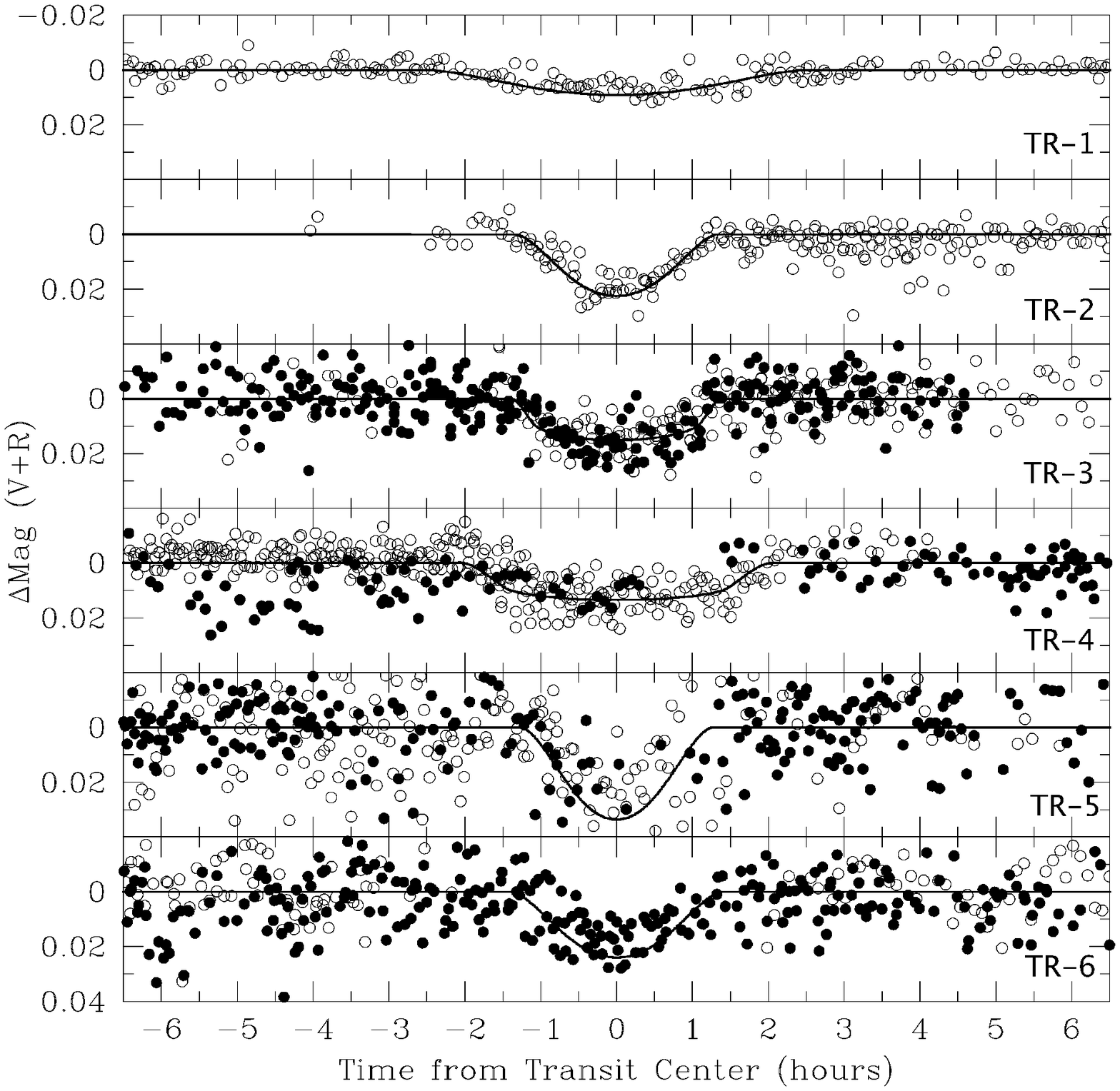} \figcaption[photplot.eps]{Light curves of the
  six transit candidates (Lupus-TR-1 at the top, in order to
  Lupus-TR-6 at the bottom).  The open circles
  represent 2005 data and the filled circles represent data from 2006.
   The best fitting model transit is overplotted as a solid line.
  Lupus-TR-1 and TR-2 are both close to the saturation limit, and in
  2006 (due to the mirror being re-silvered), they were saturated in
  most frames, thus new data were not obtained.\label{candsplot}}

\plotone{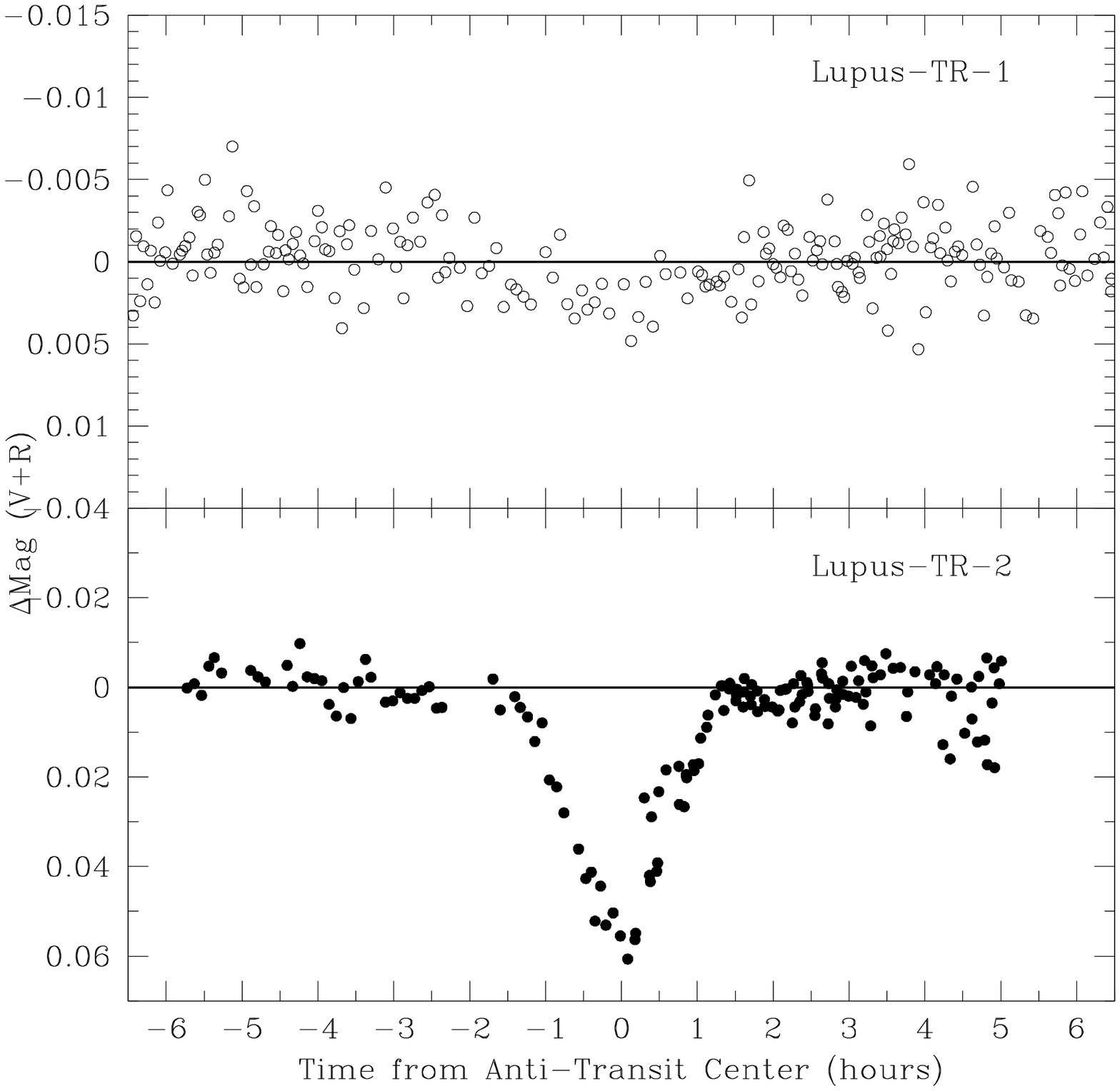} \figcaption[sec_ecl.eps]{$V+R$ photometry
  showing the secondary eclipse of Lupus-TR-1 (top) and Lupus-TR-2
  (bottom).  Again the open circles represent 2005 data and the filled
  circles represent 2006 data.  Lupus-TR-1 shows a 0.8~mmag secondary
  eclipse.  Lupus-TR-2 shows a large 6\% ``secondary eclipse'', which was
  not seen in the 2005 data.  Lupus-TR-2 is therefore a detached
  eclipsing binary. \label{TR12}}

\plotone{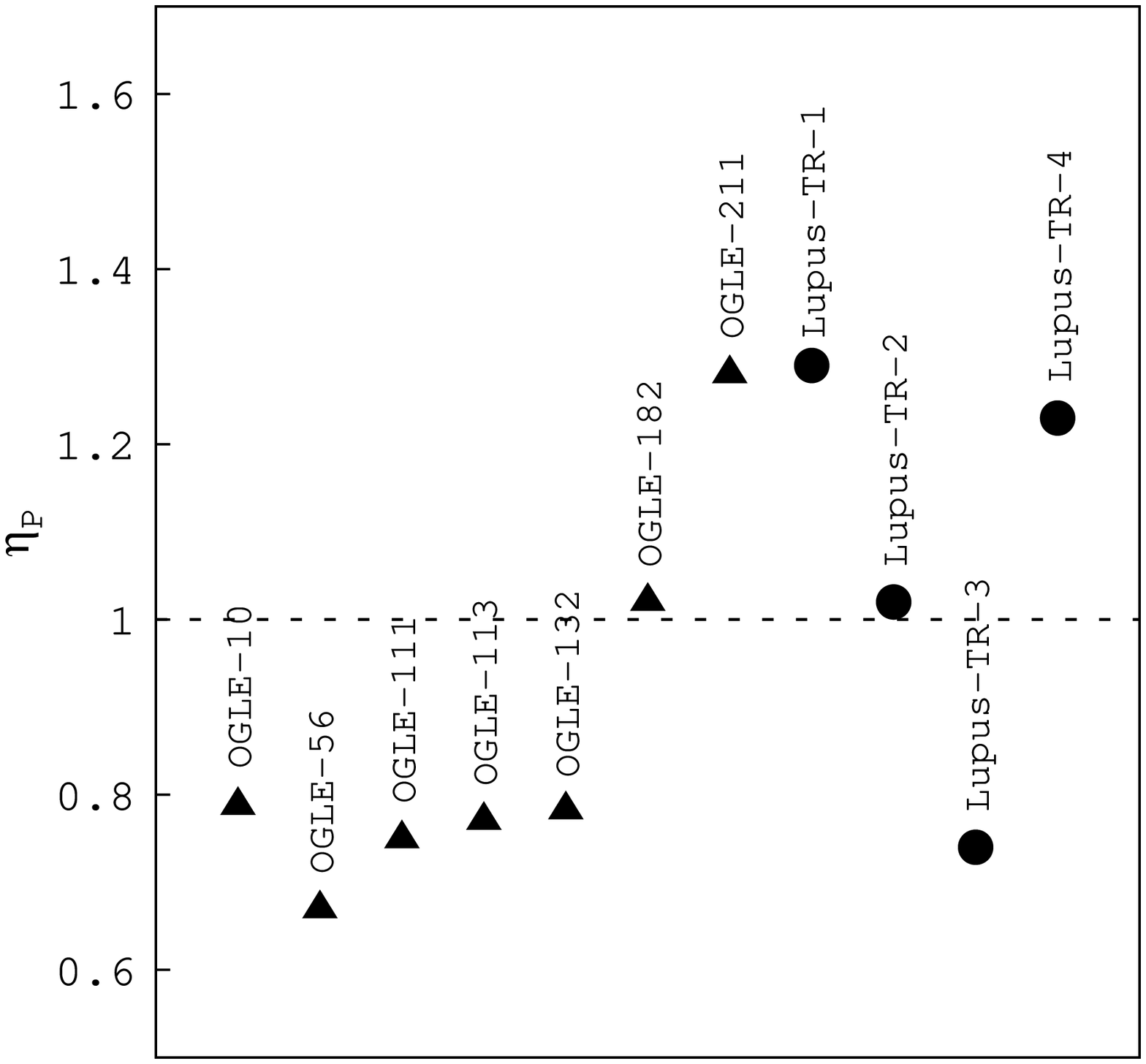} \figcaption[eta.eps]{Diagnostic number
  ($\eta_{p}$) for the Lupus candidates (circles) and, for comparative
  purposes, the OGLE planets (triangles). \label{eta}}

 \plotone{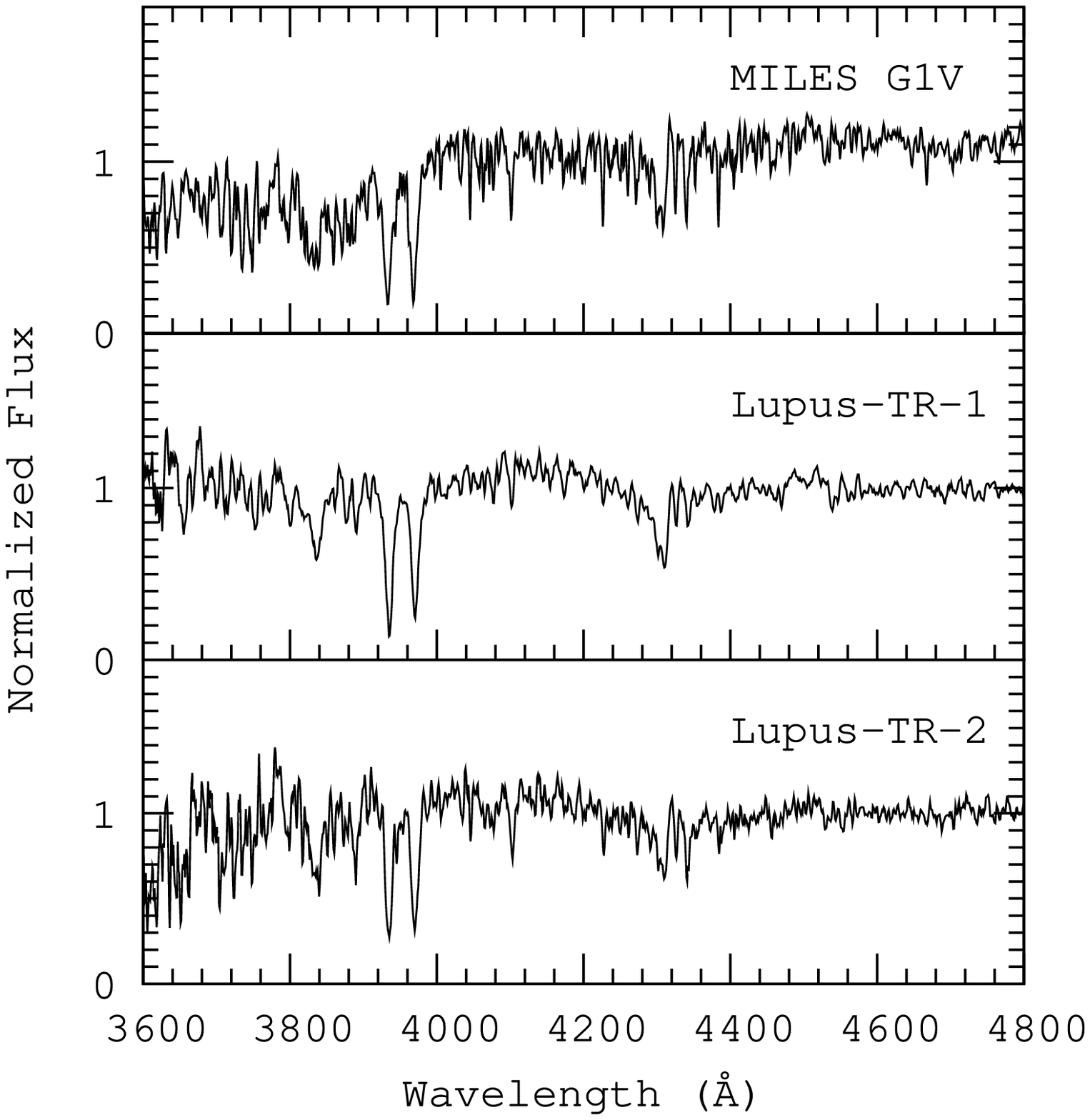} \figcaption[spec1.eps]{Spectra
   of a G1V from the MILES catalog (top), Lupus-TR-1 (middle), and
   Lupus-TR-2 (bottom).  The spectra of the Lupus candidates were taken
   using the blue arm of the DBS on the ANU 2.3m telescope at SSO.
   \label{tr12spec}}

\plotone{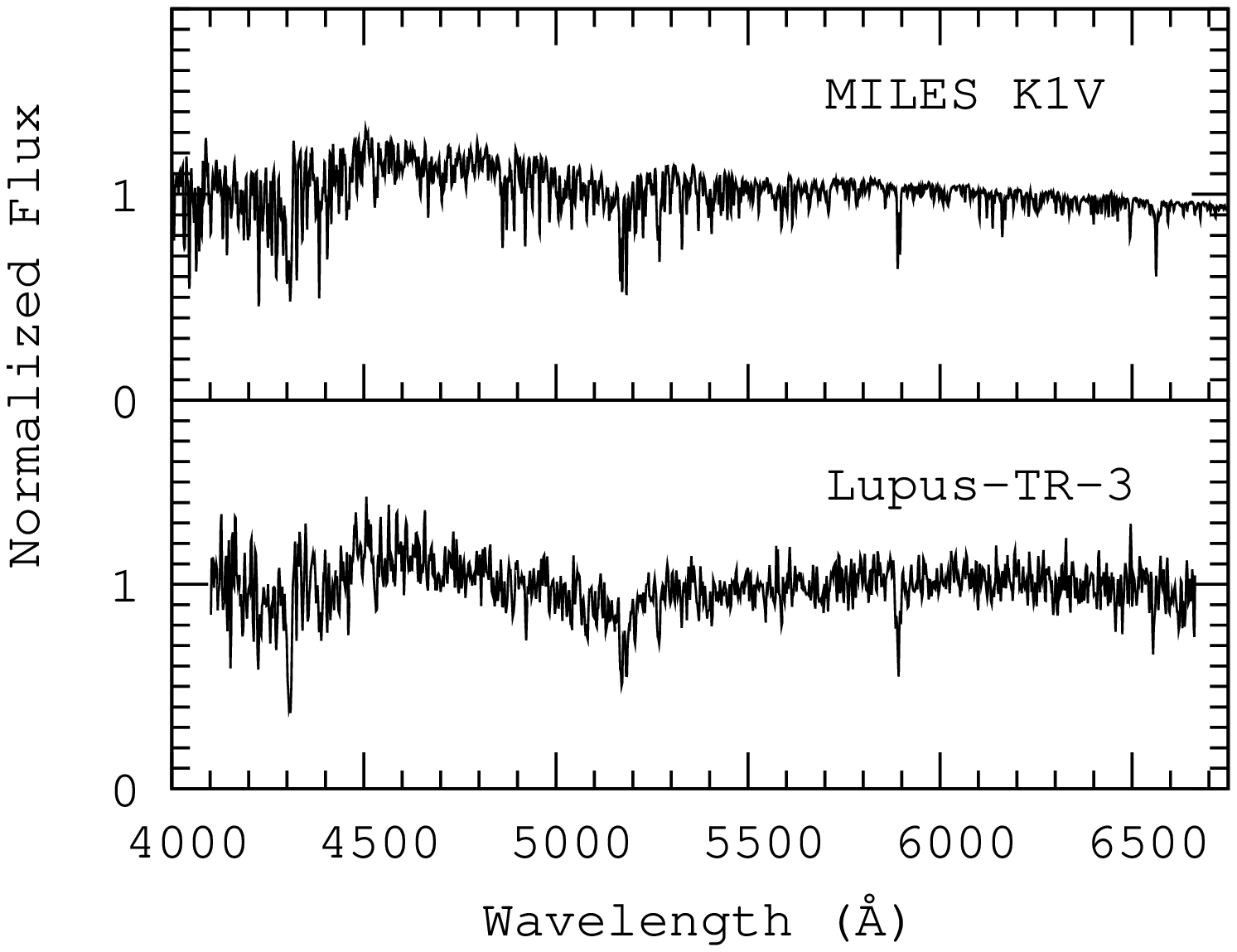} \figcaption[spec2.eps]{Spectra of a
  K1V from the MILES catalog (top) and Lupus-TR-3 (bottom).  The
  Lupus-TR-3 spectrum was taken using the blue arm of the DBS on
  the ANU 2.3m telescope at SSO. \label{tr3spec}}

\clearpage

\plotone{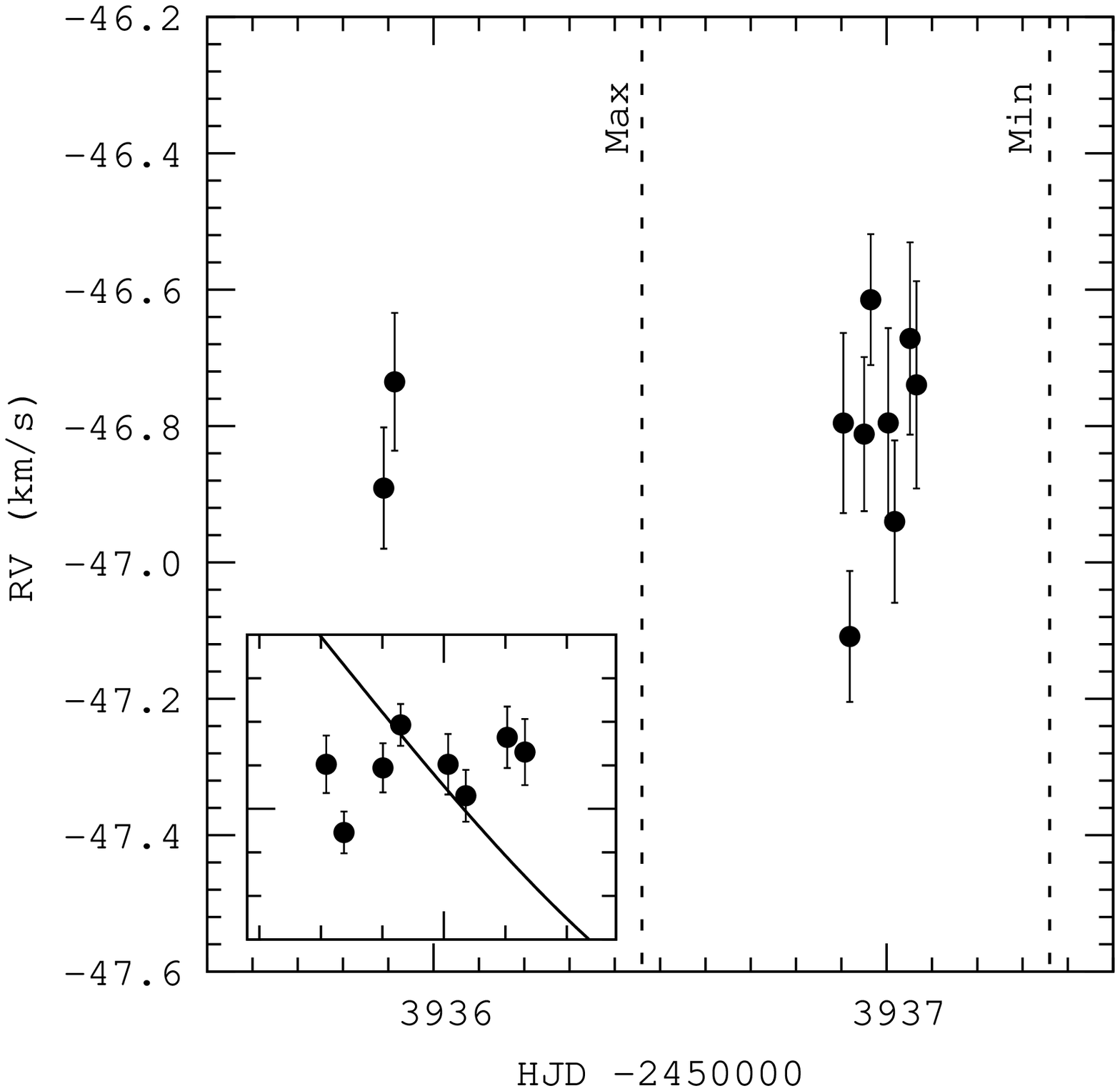} \figcaption[aat.eps]{RV measurements for Lupus-TR-1 taken using UCLES on the
  AAT. The ``Max'' and ``Min'' dashed lines indicate the time of expected maximum and
  minimum RV variation.   Inset: a magnified view of data from night
  2 and a best fitting sine curve with $K=2$~km\,s$^{-1}$ (solid line).
  The data are clearly inconsistent with such an RV variation,
  indicating Lupus-TR-1 has no binary stellar companion with a 1.8 day
  period.  It does not, however, rule out a blend scenario.\label{AAT}}

\plotone{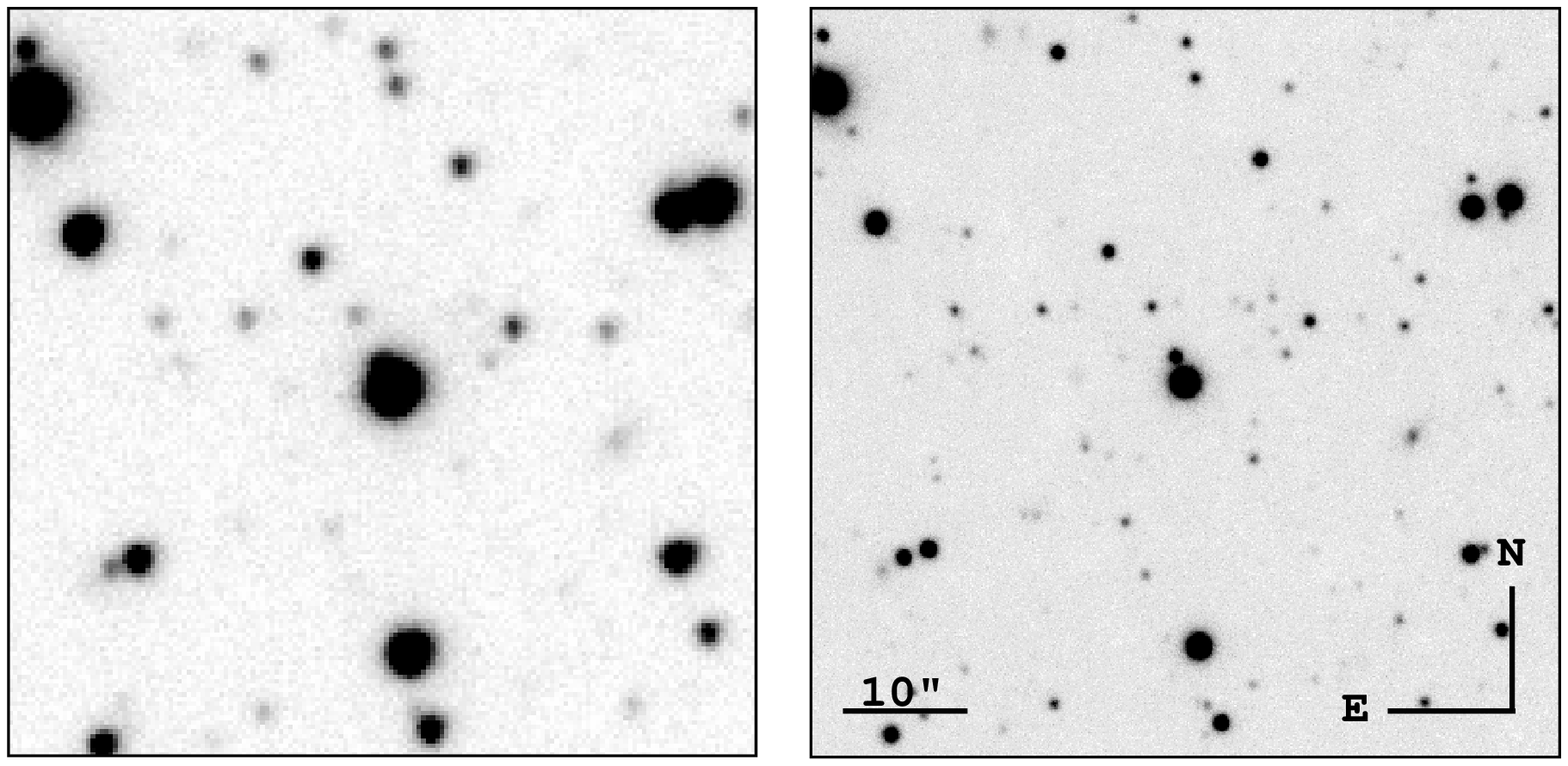} \figcaption[panic.eps]{Left: image of
  Lupus-TR-1 (central star in frame) from WFI on the ANU 1~m telescope at
  SSO (FWHM=1.2 $\arcsec$). Right: image of Lupus-TR-1 taken with PANIC on
  Magellan I (Baade) (FWHM=0.65 $\arcsec$).  The scale and
  orientation for both images are set out in the right panel.  The neighbor is
  clearly resolved in the PANIC image NNE of Lupus-TR-1 with a
  separation of 2.3$\arcsec$, and is very likely the blended source of
  the transit feature. \label{PANIC}}

\clearpage

\begin{deluxetable}{ll}
\tablecaption{Properties of the Lupus Transit Survey} 
\tablehead{
  \colhead{Property} & 
  \colhead{Lupus Transit Survey}
}
\startdata 
Telescope & The ANU 40-inch telescope\\ 
Telescope Aperture & 1~m\\ 
Telescope Site & SSO, Australia\\ 
Site Longitude & 149.1$^{\circ}$ East\\ 
Site Latitude & 31.3$^{\circ}$ South\\ 
Site Altitude & 1150~m\\ 
Detector & The WFI\\ 
Detector Size (pixels) & 8192 pixels $\times$ 8192 pixels\\
Array configuration & Eight 2k$\times$4k Lincoln Lab thinned CCDs\\ 
Pixel size (physical) & 15 $\mu$m\\ 
Pixel size (on sky) & 0.375 \arcsec\\
Field of view & 52\arcmin $\times$ 52\arcmin\\ 
Readout time & $\approx$ 50~s 
\enddata
\label{details}
\end{deluxetable}

\clearpage

\begin{deluxetable}{llllll}
\tablecaption{Candidate positions, magnitudes, colors, and spectral type} 
\tablehead{
  \colhead{ID} & 
  \colhead{RA} & 
  \colhead{DEC} &
  \colhead{V} & 
  \colhead{V$-$I} &
  \colhead{Spectral}
\\
  \colhead{}&
  \colhead{(J2000.0)}&
  \colhead{(J2000.0)}&
  \colhead{}&
  \colhead{} &
  \colhead{Type}
  }
\startdata
Lupus-TR-1 & 15:30:51.6 & -42:31:21.9 & 14.63 & 0.67 & G1V\\ 
Lupus-TR-2 & 15:30:57.3 & -42:59:02.4 & 14.78 & 0.63 & G1V\\ 
Lupus-TR-3 & 15:30:18.7 & -42:58:41.5 & 17.40 & 0.86 & K1V\\ 
Lupus-TR-4 & 15:31:04.8 & -42:57:10.3 & 16.33 & 0.77 &  \nodata \\ 
\enddata
\label{cand_data}
\end{deluxetable}

\clearpage

\begin{deluxetable}{lllllll}
\tablecaption{Candidate Properties}
\tablehead{
  \colhead{ID} & 
  \colhead{Period} & 
  \colhead{Depth ($d$)} & 
  \colhead{Duration } & 
  \colhead{$T_{C}$}&
  \colhead{$\eta$}&
  \colhead{S/N}
  \\
  \colhead{}&
  \colhead{(days)}&
  \colhead{(mag)}&
  \colhead{$T_{14}$(hours)}&
  \colhead{(HDJ-2450000)}&
  \colhead{}&
  \colhead{}
  }
\startdata
Lupus-TR-1 & 1.81820& 0.007$\pm$0.006 & 4.98$\pm$0.52 &
3525.998$\pm$0.015 & 1.29 & 20.9 \\ 
Lupus-TR-2 & 3.07070& 0.020$\pm$0.008 & 2.67$\pm$0.10 &
3530.057$\pm$0.072& 1.02 & 26.3  \\ 
Lupus-TR-3 & 3.91405& 0.016$\pm$0.004 & 2.65$\pm$0.07 &
3887.092$\pm$0.002 & 0.74 & 24.2 \\ 
Lupus-TR-4 & 2.05398& 0.013$\pm$0.004 & 3.91$\pm$0.12 &
3904.987$\pm$0.022 & 1.23 & 23.2\\
\enddata
\label{cand_properties}
\end{deluxetable}

\clearpage

\begin{deluxetable}{rrr}
\tablecaption{AAT RV Measurements for Lupus-TR-1}
\tablehead{
  \colhead{HJD} & 
  \colhead{RV} & 
  \colhead{$\pm1\sigma$} \\
  \colhead{(-2450000}) & 
  \colhead{(km s$^{-1}$)} &
  \colhead{(km s$^{-1}$)} 
} 
\startdata
3935.890 & -46.89 & 0.09 \\ 
3935.914 & -46.74 & 0.10 \\ 
3936.904 & -46.80 & 0.13 \\ 
3936.919 & -47.11 & 0.10 \\ 
3936.950 & -46.81 & 0.11 \\ 
3936.965 & -46.61 & 0.10 \\ 
3937.003 & -46.80 & 0.14 \\ 
3937.018 & -46.94 & 0.12 \\ 
3937.052 & -46.67 & 0.14 \\ 
3937.066 & -46.74 & 0.15 
\enddata
\label{AAT_RV}
\end{deluxetable}


\begin{thebibliography}{}

\bibitem[Alard \& Lupton(1998)]{alard98} Alard, C., \& Lupton, R.~H.\
  1998, \apj, 503, 325
\bibitem[Alonso et al.(2004)]{alonso04} Alonso, R., et al.\ 2004,
  \apjl, 613, L153
\bibitem[Bakos et al.(2007)]{bakos07} Bakos, G.~\'A., et al.\ 2007,
  \apj, 656, 552
\bibitem[Bayliss \& Sackett(2007)]{bayliss07} Bayliss, D.~D.~R., \&
  Sackett, P.~D. \ 2007, in ASP Conf. Ser. 366, Transiting Extrasolar
  Planets Workshop ed C. Afonso, D. Weldrake, \& Th. Henning (San
  Francisco, CA: ASP), 320
\bibitem[Bayliss et al.(2008)]{bayliss08} Bayliss, D. D. R., Sackett,
  P. D., \& Weldrake, D. T. F \ 2008, in IAU Symposium 253, Transiting
  Planets, in press (arXiv:0807.0469v1)
\bibitem[Charbonneau et al.(2005)]{charbonneau05} Charbonneau, D., et
  al.\ 2005, \apj, 626, 523
\bibitem[Collier Cameron et al.(2007)]{cameron07} Collier Cameron, A.,
  et al.\ 2007, \mnras, 375, 951
\bibitem[Deming et al.(2005)]{deming05} Deming, D., Seager, S.,
  Richardson, L.~J., \& Harrington, J.\ 2005, \nat, 434, 740
\bibitem[Hartman et al.(2004)]{H2004} Hartman, J.~D., Bakos, G.,
  Stanek, K.~Z., \& Noyes, R.~W.\ 2004, \aj, 128, 1761
\bibitem [Kov\'acs, Zucker, \& Mazeh(2002)]{kovacs02} Kov\'acs, G.,
  Zucker, S., \& Mazeh, T. \ 2002, \aap, 391, 369
\bibitem[Landolt(1992)]{Lan1992} Landolt, A.~U.\ 1992, \aj, 104, 340
\bibitem[Mandel \& Agol(2002)]{MA2002} Mandel, K., \& Agol, E.\ 2002,
  \apjl, 580, L171
\bibitem [McCullough et al.(2006)]{mccullough06} McCullough, P.~R., et
  al.\ 2006, \apj, 648, 1228
\bibitem [Pont, Zucker, \& Queloz(2006)]{pont06} Pont, F., Zucker, S.,
  \& Queloz, D.\ 2006, \mnras, 373, 231
\bibitem [Pont et al.(2008)]{pont08} Pont, F., et al.\ 2008, \aap, 487, 749
\bibitem [Sackett et al.(2008)]{sackett08} Sackett, P.~D., Gillon, M.,
Bayliss, D.~D.~R., Weldrake, D.~T.~F., Tingley, B.~W. \ 2008, in IAU
Symposium 253, Transiting Planets, in press
\bibitem[S{\'a}nchez-Bl{\'a}zquez et al.(2004)]{sanchez04}
  S{\'a}nchez-Bl{\'a}zquez, P., Peletier, R. F., Jim{\'e}nez-Vicente,
  J., Cardiel, N., Cenarro, A. J., Falc{\'o}n-Barroso, J., Gorgas, J.,
  Selam, S., \& Vazdekis, A.\ 2006, \mnras, 371, 703
\bibitem[Schlegel et al.(1998)]{schlegel98} Schlegel, D.~J.,
  Finkbeiner, D.~P., \& Davis, M.\ 1998, \apj, 500, 525
\bibitem[Stetson(2000)]{Stet2000} Stetson, P.~B.\ 2000, \pasp, 112, 925
\bibitem[Tamuz et al.(2005)]{tamuz05} Tamuz, O., Mazeh, T., \& Zucker,
  S.\ 2005, \mnras, 356, 1466
\bibitem[Tinetti et al.(2007)]{tinetti07} Tinetti, G., et al.\ 2007,
  \nat, 448, 169
\bibitem[Tingley \& Sackett(2005)]{tingley05} Tingley, B., \& Sackett,
  P.~D.\ 2005, \apj, 627, 1011
\bibitem[Udalski et al.(2002)]{udalski02} Udalski, A., et al.\ 2002,
  \actaa, 52, 1
\bibitem[Udalski(2007)]{udalski07} Udalski, A.\ 2007, in ASP
 Conf. Ser. 366, Transiting Extrasolar Planets Workshop ed C. Afonso,
 D. Weldrake, \& Th. Henning (San Francisco, CA: ASP), 51
\bibitem[Vidal-Madjar et al.(2004)]{vidal04} Vidal-Madjar, A., et al.\
  2004, \apjl, 604, L69
\bibitem[Weldrake \& Bayliss(2008)]{weldrake08} Weldrake, D.~T.~F., \&
  Bayliss, D.~D.~R.\ 2008, \aj, 135, 649
\bibitem[Weldrake et al.(2008)]{weldrake08b} Weldrake, D.~T.~F.,
  Bayliss, D.~D.~R., Sackett, P.~D., Tingley, B., Gillon, M., \&
  Setiawan, J.\ 2008, \apjl, 675, L37
\bibitem[Weldrake et al.(2007)]{weldrake07} Weldrake, D.~T.~F.,
  Sackett, P.~D., \& Bridges, T.~J.\ 2007, \aj, 133, 1447
\bibitem[Weldrake et al.(2005)]{weldrake05} Weldrake, D.~T.~F.,
  Sackett, P.~D., Bridges, T.~J., \& Freeman, K.~C.\ 2005, \apj, 620,
  1043
\bibitem[Weldrake \& Sackett(2005)]{weldrake05b} Weldrake, D.~T.~F.,
  \& Sackett, P.~D.\ 2005, \apj, 620, 1033
\bibitem[Wozniak(2000)]{wozniak00} Wozniak, P.~R.\ 2000, \actaa, 50,
  421
\end{thebibliography}
\end{document}